\documentclass[preprint]{aastex}

\shortauthors{Pike & Kavelaars}
\begin{document}

\title{ON A POSSIBLE SIZE/COLOR RELATIONSHIP IN THE KUIPER BELT}
\slugcomment{accepted to The Astronomical Journal July 18 2013}
\shorttitle{Kuiper belt colors}
\author{R. E. Pike\altaffilmark{1,2}}
\altaffiltext{1}{Department of Physics and Astronomy, University of Victoria, Victoria, BC, V8W 3P6l, Canada }
\email{repike@uvic.ca}

\author{J.  J.  Kavelaars\altaffilmark{2,1}}
\altaffiltext{2}{National Research Council of Canada, 5071 West Saanich Road, Victoria, BC, V9E 2E7, Canada}

\begin{abstract}
Color measurements and albedo distributions introduce non-intuitive observational biases in size-color relationships among Kuiper Belt Objects (KBOs) that cannot be disentangled without a well characterized sample population with systematic photometry.  
Peixinho et al. report that the form of the KBO color distribution varies with absolute magnitude, $H$.  
However, Tegler et al. find that KBO color distributions are a property of object classification.  
We construct synthetic models of observed KBO colors based on two $B-R$ color distribution scenarios: color distribution dependent on $H$ magnitude ($H$-Model) and color distribution based on object classification (Class-Model). 
These synthetic $B-R$ color distributions were modified to account for observational flux biases.
We compare our synthetic $B-R$ distributions to the observed `Hot' and `Cold' detected objects from the Canada-France Ecliptic Plane Survey and the Meudon Multicolor Survey.  
For both surveys, the Hot population color distribution rejects the $H$-Model, but is well described by the Class-Model.
The Cold objects reject the $H$-Model, but the Class-Model (while not statistically rejected) also does not provide a compelling match for data.
Although we formally reject models where the structure of the color distribution is a strong function of $H$ magnitude, we also do not find that a simple dependence of color distribution on orbit classification is sufficient to describe the color distribution of classical KBOs.

\end{abstract}


\section{Introduction}

The first color measurements of KBOs found that they exhibited a surprising diversity of colors.
If KBOs were formed in situ, they would have nearly identical compositional properties and weathering histories because the region between 34 AU and $\sim$60 AU is characterized by a single set of stable volatile materials \citep{brown}.
Color measurements of KBOs, however, show a range from very red ($B-R\sim2.0$) to neutral or blueish ($B-R\sim1.0$), compared to a solar $B-R$ of 1.03.  
There is little consensus on the origin of the KBO color heterogeneity. 

The effects of collisional processing and weathering on KBO colors have been considered.
\cite{LuuJewitt} performed numerical simulations which incorporated a reddening component from irradiation and a collisional erosion process which exposed neutral material on the objects' surfaces.  
This model successfully reproduces the range of colors observed in KBOs, but also predicts objects with large photometric light curves due to surface color variations and partial resurfacing from impacts.  
This was later found to be inconsistent with the small magnitude variations for KBOs, regardless of $B-R$ color \citep{jewitt}.  
The model also does not predict the correlated colors of binary KBOs \citep{benecchi}.  
A balance of collisional resurfacing and weathering processes was found to be incompatible with the observed properties of KBOs.

Space weathering may still have a measurable impact on the apparent colors of KBOs.
\cite{kanuchova} used a two component model of KBO composition and investigated the effects of weathering due to solar wind ions.
Their lab experiments showed that irradiation caused reddening in the carbon-bearing material, but further exposure resulted in a darkening and flattening of the spectra.
Their model incorporated a component with a flat spectrum and a silicate material with a carbon-bearing layer.
The varying combination of these components with different amounts of irradiation reproduces the range of colors seen in the outer solar system.
If the colors of KBOs are a result of solar wind irradiation, the entire evolutionary history, as well as original composition, would affect the apparent color of KBOs.
 
Different KBO colors may primarily reflect compositional variation during formation.
\cite{brown} argue that when objects form outside of approximately 20 AU, methanol ice (CH$_{4}$O) is stable and expected to be the most abundant component.  
The irradiation of this ice will result in a red object with a high albedo \citep{brunetto}.  
Inward of approximately 20 AU, the surfaces of objects are composed of water (H$_{2}$O), carbon dioxide (CO$_{2}$), and small amounts of hydrogen sulfide (H$_{2}$S), creating a dark, neutral surface color when irradiated \citep{closeice1, closeice2}.  
Transport of this assortment of KBOs from their different source regions to their current locations in the Kuiper Belt may explain the distribution of neutral and red KBOs observed.

At least two scenarios of KBO color distribution have been proposed.
\cite{tegler03} found that the Cold classical KBOs (cKBOs) were red while the Hot cKBOs were gray.  
Hot and Cold objects were differentiated using perihelion, $q$, and inclination, $i$, and they found two different unimodal color distributions.  
Different color distributions for different dynamical classifications suggests that the Hot and Cold cKBOs have different histories, perhaps different formation locations.
A recent paper by \cite{peixinho} performed a detailed statistical analysis of KBO colors based on $H$ magnitude, a proxy for size, searching for a correlation with $B-R$ color distributions.
Data were compiled from previous surveys, and 253 objects were included in the analysis.  
The colors of small objects ($H >$ 6.8) were found to be well described by a bimodal distribution.  
The larger objects ($H < 5.0$) were also found to follow a bimodal distribution, but with slightly different peak values.
The intermediate objects were found to be unimodal, covering the full range of $B-R$ values seen in the other $H$ ranges.  
If this $H$ magnitude dependency is true, new models will be required to understand the physical processes that create this effect.
Resolving the nature of the color diversity is a necessary step in understanding the global evolution of the Kuiper Belt.

KBO discoveries and followup observations have had limited success because of the challenges presented by several observational biases.  
The discovery of KBOs in a survey results in predicted observational arcs.  
Depending on the prediction algorithm, assumptions are made about the objects' orbits, such as an energy constraint, a circular orbit, or the detection location (perihelion or aphelion).
These different assumptions have different recovery rates for different dynamical classes of object \citep{jones}.  
The size of the recovery field also has a significant impact on the likelihood of successfully detecting the KBO during followup observations.
These challenges are reflected in the Minor Planet Center database, which often includes objects that were ``lost'' or have ambiguous orbital parameters.  
KBOs with poorly predicted orbits may even have incorrect orbital classifications.
 It can require years of observations to constrain an orbit accurately enough to determine if an object is in a resonance \citep{gladman}.  
The discovery wavelength can also have an impact on the colors of discovered objects; redder objects are easier to find in an $R$ band survey, but extremely red objects are difficult to measure in a range of photometric colors.  
The effect of these biases becomes more difficult to quantify when data are combined from multiple surveys.

We investigated the $B-R$ color distributions of cKBOs using two sample populations.
We used a sample from the Canada-France Ecliptic Plane Survey \citep[CFEPS;][]{cfepsSS}, with a known discovery method, and a sample from the Meudon Multicolor Survey (2MS) photometric measurements \citep{2ms}, which has unknown selection effects.
The KBOs were classified into Hot and Cold populations in two different ways: based on their $q$ and semi-major axis, $a$, as well as by an $i$ cut.
The $a$ and $q$ cuts appear to provide a better division between the Hot and Cold cKBO populations in the CFEPS L7 model \citep{cfepsSS}.
We created synthetic cKBO populations with color distributions based on object classifications and $H$ magnitudes and attempted to account for the biases introduced by photometric followup observations.
The cKBO model populations were compared to the real cKBO samples using the Anderson-Darling (AD) statistical test.
We find that an $H$ magnitude based color distribution is statistically rejected by the observations, but color distributions purely based on dynamical classification appear to be insufficient to describe the color distribution of cKBOs.

\section{Sample populations}

\begin{figure}[h!]
\begin{center}
\includegraphics[width=1\textwidth]{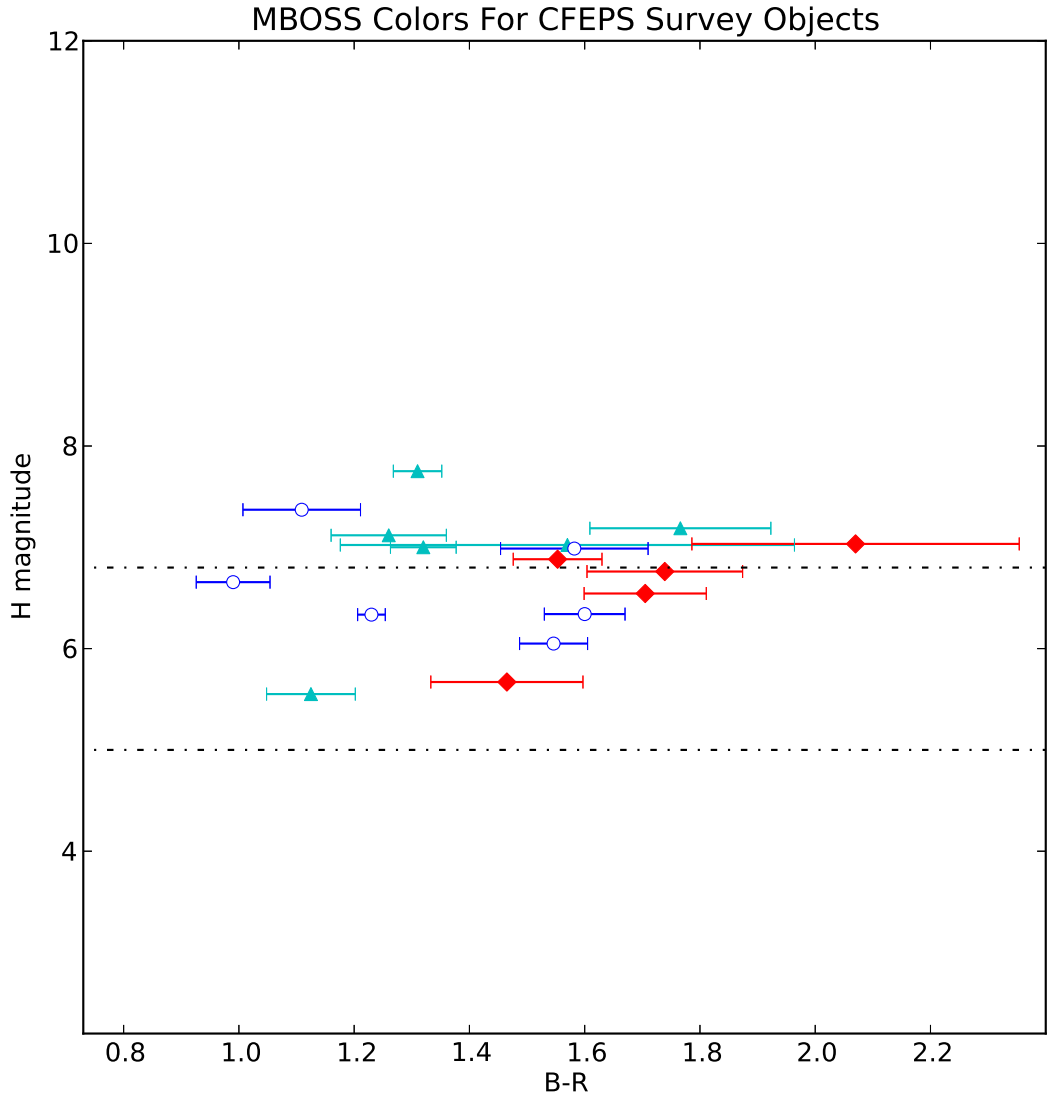}
\caption{CFEPS KBO detections are cross referenced with the MBOSS color database.  Hot objects are marked by triangles, Cold objects are marked with diamonds, and Resonant objects are shown as open circles.  The classical objects have been classified using their $q$ and $a$ values, as described in the text.  The horizontal lines indicate the regions of different $B-R$ color distribution according to the $H$-Model.}
\label{cfeps_obj}
\end{center}
\end{figure}

\begin{figure}[h!]
\begin{center}
\includegraphics[width=1\textwidth]{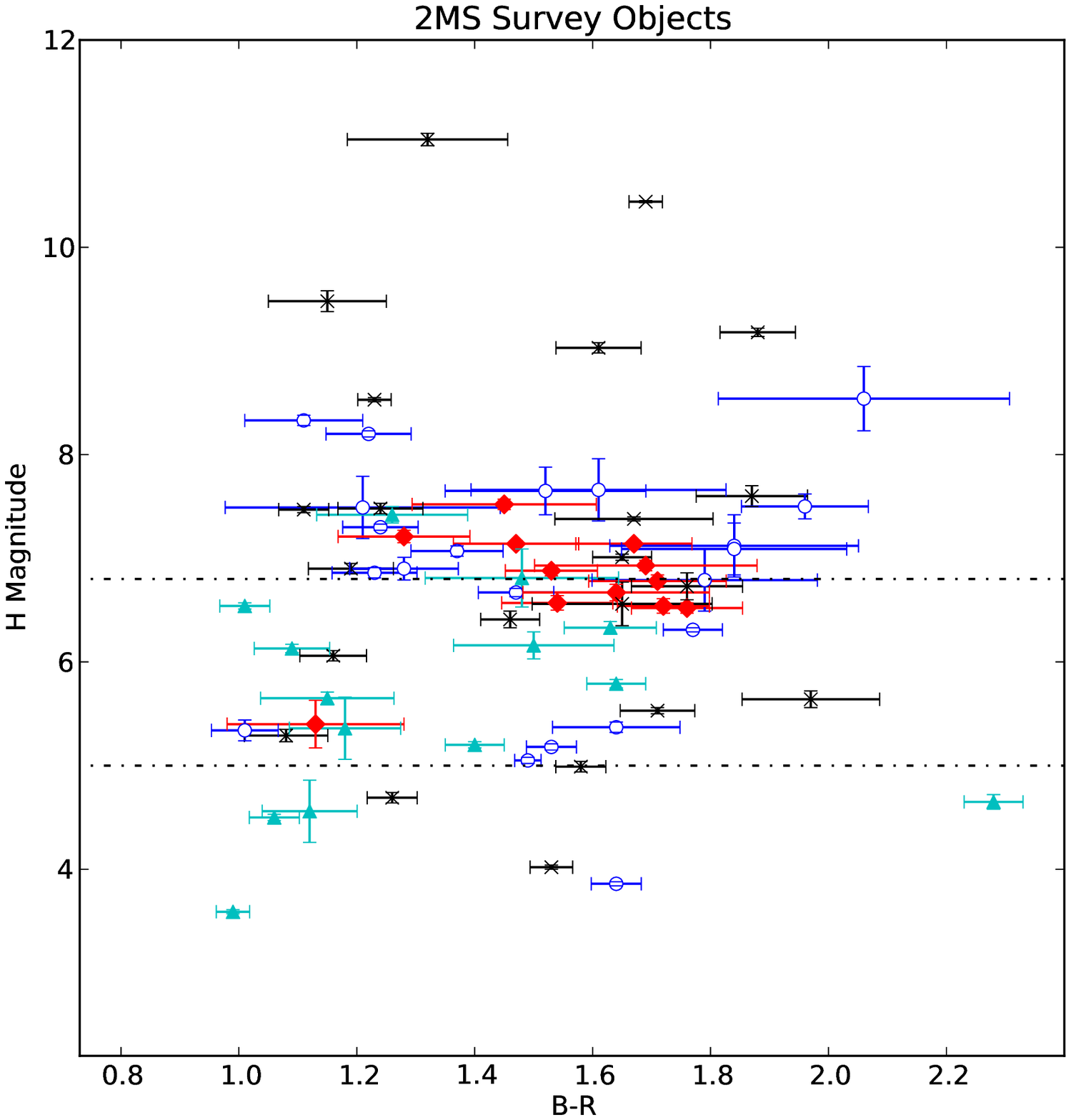}
\caption{The Meudon Multicolor Survey objects are indicated by the same symbols as in Figure \ref{cfeps_obj}.  Other object types, such as Centaurs, are marked with an `$\times$'.  The Cold objects primarily cluster around $B-R$ of 1.6.  The Hot object at a $H$ magnitude of 4.6 and $B-R$ of 2.3 appears to be an outlier and was excluded from the analysis.  The horizontal lines indicate the regions of different $B-R$ color distribution according to the $H$-Model.}
\label{2ms_obj}
\end{center}
\end{figure}

To distinguish between competing models for KBO color distributions, we compare published KBO color measurements with model based distributions.  
We consider two independent samples of KBO colors; a sample derived from the  CFEPS L7 release  \citep{cfepsSS}, and one based on the 2MS data  \citep{2ms}.
The CFEPS L7 sample was cross referenced with the MBOSS database of photometric KBO observations \citep{mboss} to determine the photometric properties of the sample.  
Selecting only the classical CFEPS L7 objects that have high-quality photometric measurements in the MBOSS database results in a sample of 8 objects, shown in Figure \ref{cfeps_obj}.   
The 2MS sample is a survey of photometric observations, and those values were used directly.  
The 2MS sample is considerably larger (21 objects), however the selection criterion of the data are unknown.
The $B-R$ colors of the 2MS sample are shown in Figure \ref{2ms_obj}. 
Each of these KBO color sample populations is compared to a set of synthetic detections. 

Two sets of criterion were used to classify the sample objects into Hot and Cold subpopulations.
One classification scheme is based on their $q$ and $a$ values and the other is based on a cut in $i$.
For the $q$ criterion,  cKBOs with $35 < q < 40$ AU are placed in the Hot subpopulation and those with $q > 40$ AU and $ 42.5 < a < 47.2$ AU in the Cold. 
For the $i$ criterion, Hot cKBOs  have $i > 7^{\circ}$ and Cold ones have $i< 4^{\circ}$.  
A gap in the inclination cuts for this criterion minimizes possible contamination between the subpopulations.
All orbital parameters used in these classifications were taken from the Minor Planet Center database.

\section{Methods}

We created model KBO populations and color distributions in order to compare the sample populations with the color distribution models.
Different synthetic KBO populations were created for the CFEPS and 2MS observations.
The synthetic objects were assigned colors based on color distribution models to create the $H$-Model and the Class-Model.
We also included the effect of photometric biases, resulting from observing the intrinsic population, into our synthetic detection models.
These models, incorporating KBO orbital characteristics, $B-R$ color, and measurement bias, were used to test the consistency of the observations with color distribution models.

\subsection{Model Populations}

The L7 synthetic model provides a reasonable representation of the intrinsic Kuiper Belt \citep{cfepsSS}.  
Using the CFEPS L7 characterization and the input L7 model, we build a simulated sample of detected CFEPS KBOs.
For these simulated detections, the orbital elements, $H$ magnitudes, and dynamical classifications of the objects are known.
The expected flux and $H$ magnitude measurement uncertainties are modeled.
This process allows us to understand the photometric biases present in the CFEPS cKBO color samples. 

For the 2MS sample, a bootstrapped comparison population was derived.
The selection biases introduced in the 2MS discovery methods are not known. 
Because the objects were not discovered in a systematic survey, the 2MS objects did not have a representative distribution of $H$ magnitudes.
The synthetic detections were created by resampling the actual objects from within their $H$ magnitude uncertainties.
This generated a large model population for comparison, which could be assigned colors based on $H$ magnitude and orbital parameters representative of the 2MS data.

The cKBOs were selected from the synthetic populations, and classified into Hot and Cold objects.
(Resonant objects are highly sensitive to survey characteristics and are excluded from our analysis.)
Different classification schemes for the CFEPS and 2MS synthetic detection models were required because of the different methods used to create the models.
For the CFEPS synthetic objects, we used the known model classifications \citep{cfepsSS}.
The 2MS synthetic objects were classified in the same way as the 2MS data, based on their $q$, $a$, and $i$ values.
These theoretical model populations provide a good match to the Kuiper Belt orbital distribution as sampled by the data and were used as our Hot and Cold comparison populations for modeling the various $B-R$ color distributions.

\subsection{$H$ Magnitude Color Model}
The $B-R$ color distribution of the $H$-Model depends on the $H$ magnitude of the objects, as described in \cite{peixinho}.  
Three different $H$ magnitude ranges were identified.
\citet{peixinho} determined that objects with $H > 6.8$ and the objects with $H < 5.0$ were bimodal, while the intermediate $H$ magnitude objects were found to be unimodal.  
The bimodal low and high $H$ magnitude populations also had different $B-R$ peak values defining their distribution.  
For the $H > 6.8$ objects we used a bimodal $B-R$ color distribution with peaks at 1.2 and 1.8, each represented by a gaussian with a standard deviation of 0.125 mag.  
The intermediate $H$ objects were modeled as a single broad distribution centered at $B-R$ of 1.6 with a standard deviation of 0.33 mag.  
The $H < 5$ objects had peak $B-R$ values of 1.05 and 1.6 with standard deviations of 0.1 mag.  
$B-R$ colors were generated randomly from these distributions, and assigned to the synthetic objects in the model populations based on their $H$ magnitudes to create the $H$-Model.

\subsection{Classification Color Model}
\begin{figure}[h!]
\begin{center}
\includegraphics[width=1\textwidth]{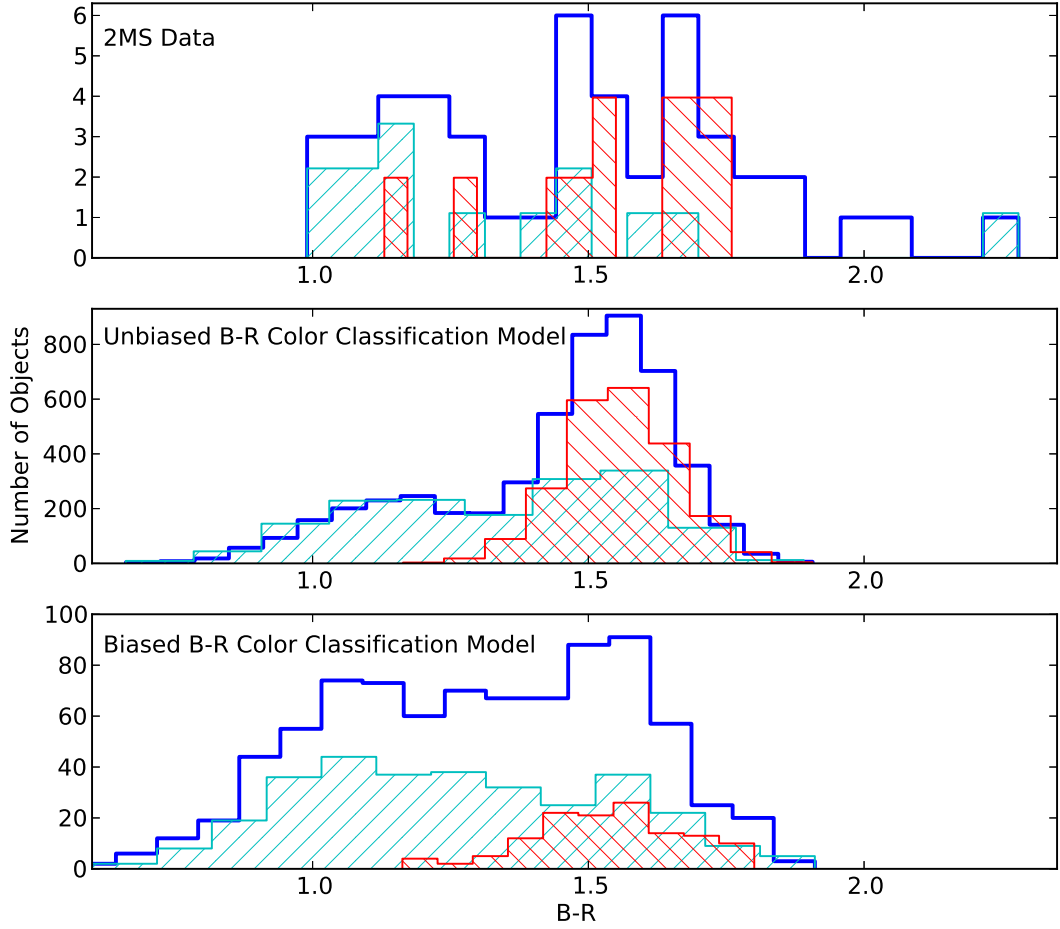}
\caption{Panel 1: The 2MS object $B-R$ magnitudes.  The complete sample (thick line) includes only the Hot (thin line with `/'), Cold (thin line with `\textbackslash{}'), and Resonant objects (not plotted separately).  The very red objects that are not well described by the model include an excluded Hot object and Resonant objects not considered in this analysis.  Panel 2: The Class-Model with Hot and Cold objects.  Panel 3: The Class-Model with color measurement bias.}
\label{hist}
\end{center}
\end{figure}

The second $B-R$ color model assumes that the color distribution of the KBOs is based on the objects' dynamical classification.  
We used a color distribution with a red peak at $B-R$ of 1.55 with a standard deviation of 0.1, and a blue peak at $B-R$ of 1.15 with a standard deviation of 0.12 mag. 
These parameters were selected to match the color distribution of 2MS objects in Figure \ref{hist}.
The 2MS data have two peaks, however, the data in Figures \ref{cfeps_obj} and \ref{2ms_obj} suggest that the Cold population lacks a blue component.
We modeled the Cold objects' color distribution with a unimodal red peak, and for the Hot objects we sampled the bimodal distribution.
The distribution of $B-R$ colors for the 2MS sample and the classification based color models are shown in the histogram in Figure \ref{hist}.  
The $B-R$ colors were assigned to the synthetic objects in the model populations based on their classifications to create the Class-Model.

\subsection{Color Magnitude Bias Model}
In addition to a survey detection bias, the precision of the measured $B-R$ magnitudes will introduce an observational bias because of the brightness of the target objects.  
Color surveys of KBOs generally select objects that appear bright enough in $R$ band for photometry.  
However, for two equally bright objects in $R$, redder objects have higher $B$ magnitudes, and are therefore more difficult to measure.  
The 2MS data were used to derive a relationship between larger magnitude uncertainties and increasing $R$ magnitude, and these model uncertainties were then applied to the Class-Model and $H$-Model synthetic colors.   
A $\sim$0.4 mag separation of the two color peaks in the model suggests that photometric precision is required to successfully detect the bimodality of the intrinsic population.
The KBOs with $\delta(B-R) > 0.15$ were considered un-measurable due to their large photometric uncertainty and removed from both of the color models.
The remaining objects were assigned magnitude values within their error bars from the $R$ magnitude error model.  
These observable objects and colors with photometric error included were used to create new color models for both the $H$-Model and Class-Model, and are referred to as the color measurement biased (CMB) models.

Once the photometric uncertainty cut is applied to the model, the new CMB models show a clear correlation between classification and $H$ magnitude, for both the $H$ and Class -Model samples.  
The different physical locations of the different classes of objects make larger-$H$ objects in some classes easier to detect than others.  
This is evident even in the 2MS data in Figure \ref{2ms_obj}, where the larger-$H$ objects are all centaurs.  
In the models, the faintest objects predicted, with $H$ magnitudes greater than 7.5, are almost entirely Resonant objects.  
A mix of Resonant and Hot objects are found between $H$ magnitudes of 6.75 to 7.5, and the detectable Cold objects almost all have $H$ magnitudes below 6.75.  
The observing biases caused by photometric detection limits and $H$ magnitude dependence make it more difficult to understand the nature of the underlying population.

\subsection{Statistical Tests}

We used the AD test to compare the samples, as in CFEPS \citep{jj}.
Similar to the Kolmogorov-Smirnov test, the AD statistic computes the maximum distance between two distributions, however, the AD statistic ($D$) is weighted.
In order to determine the significance of the AD statistic value, a bootstrapping method was used.
The significance of rejection of the null hypothesis, where the distributions are identical, was calculated by comparing the real object $D$ statistics with the input model $D$ statistics.

To mitigate the effect of precise sample values in our observational data, the AD $D$ value was calculated for photometric measurements consistent with the uncertainty associated with the data.  
For each object, the magnitude measurements were randomly assigned a value within their reported uncertainty range, and then the AD $D$ values was computed. This process was then repeated 1000 times. 
The mean and the standard deviation of the range of $D$ statistic values generated from the resampled data were compared to the $D$ statistic range found using the synthetic models.  
This gave the probability that the observations, with measurement uncertainties, could be from the synthetic color models.

\section{Results}

Tables~\ref{ADtest} and \ref{ADtestC} show the agreement of observations with the different $B-R$ models used.  
The results for the color models are given, with and without the addition of the CMB.  
Both the $H$-Model (color distributions based on $H$ magnitude) and the Class-Model (colors based on object type) were compared to the CFEPS objects and the 2MS data, using both orbital subpopulation classification criteria.  
The $B-R$ distributions of Hot and Cold objects observed were compared separately against the model Hot and Cold objects.  
The use of $q$ or $i$ classification scheme is as indicated in Tables \ref{ADtest} and \ref{ADtestC}.
The range of statistical significance given in the tables provides an indicator of the robustness of the results; large ranges are primarily a result of an insufficient number of precise $B-R$ color measurements of real objects. 
Low significance values, in bold in the table, indicate that the observed $B-R$ color distribution is unlikely to have been derived from the referenced theoretical model.

\begin{table}[h!]
\caption{Statistical Test Results: Hot cKBOs}
\label{ADtest}
\begin{center}
\begin{tabular}{ l  c  c  c  c  c  c }
\hline\hline
Model & CFEPS ($q$) & 2MS ($q$) & CFEPS ($i$) & 2MS ($i$) \\\hline
$H$-Model & 10$_{10}^{20}$ & {\bf 0.2$_{0.2}^{0.2}$ }   & 40$_{30}^{40}$   & {\bf 0.9$_{0.6}^{1}$ } \\
$H$-Model \& CMB & 20$_{20}^{40}$  & {\bf 0.7$_{0.4}^{0.9}$ }  & 70$_{50}^{90}$   & {\bf 3$_{2}^{5}$ } \\
Class-Model & 20$_{20}^{40}$  & 20$_{10}^{30}$  & 80$_{60}^{90}$   & 30$_{20}^{40} $ \\
Class-Model \& CMB & 40$_{30}^{50}$ & 80$_{60}^{90}$   & 80$_{70}^{90}$   & 90$_{70}^{90}$ \\
\hline
\end{tabular}
\end{center}
\tablenotetext{}{The numbers represent the percentage of times the Anderson-Darling D statistic for the observed sample was larger than the bootstrapped input model.
The bold values indicate the model is statistically rejected.
See text for details.}
\end{table}

\begin{table}[h!]
\caption{Statistical Test Results: Cold cKBOs}
\label{ADtestC}
\begin{center}
\begin{tabular}{ l  c  c  c  c  c  c }
\hline\hline
Model & CFEPS ($q$) & 2MS ($q$) & CFEPS ($i$) & 2MS ($i$) \\\hline
$H$-Model & 30$_{30}^{40}$&   5$_{5}^{10}$    & 40$_{30}^{60}$   & 9$_{7}^{10}$  \\
$H$-Model \& CMB & {\bf 4$_{2}^{6}$ } & {\bf 0$_{0}^{0.1}$ }  &  6$_{4}^{10}$  & {\bf 1$_{0.5}^{2}$ } \\
Class-Model & 30$_{4}^{50}$  & {\bf 3$_{0.4}^{20}$ }  & {\bf 0.7$_{0}^{10}$  } & {\bf 3$_{1}^{20}$ } \\ 
Class-Model \& CMB & 40$_{10}^{60}$ &  6$_{2}^{20}$ & 5$_{1}^{20}$   & 6$_{2}^{20}$  \\
\hline
\end{tabular}
\end{center}
\end{table}

\section{Discussion}

We have used two independent approaches to classification of cKBOs into Hot and Cold subpopulations: a $q$ based selection criterion and an $i$ based criterion.  
These two approaches classify the majority of objects into the same subpopulations, with a few sources moving between categories.  
The $q$ criterion was chosen to reflect the orbit model generation process as described in the appendix of \cite{cfepsSS} while the $i$ selection criterion is common in the literature.
We also explored various $i$ cuts and boundaries, none of which resulted in a significant variation in our results. 
The acceptance/rejection of various theoretical models was not found to be dependent on the classification criterion.

The results for the Hot cKBO population provide some insight into the underlying population distribution.
The CFEPS data is inconclusive due to the small number statistics.
However, the 2MS data appears to reject the $H$-Model, with and without the inclusion of the CMB.
We also find that while the classification scheme does not change the overall rejection or acceptance of the models, the $q$ based classification sharpens the rejection of the $H$-Model.
A classification based model is not rejected for the Hot objects, and, with the inclusion of the CMB, the Class-Model is an excellent match with the data.

The $B-R$ color distribution of Cold cKBOs have different compatibility issues when compared with the models.
The $H$-Model is only rejected with the inclusion of the CMB.
The CMB improves the acceptability of the Class-Model, however this model still does not provide a satisfying match to these data.
The lack of goodness of fit of the unimodal distribution in the Class-Model is likely due to the four KBOs with $B-R <$~1.3 (see Figure 3).
The inclusion of the color measurement bias expands the expected color range to include half of these objects, but these four color measurements have acceptable photometric uncertainties, which suggests that the unimodal idea for the Cold objects may require revision.
The Cold KBOs have a steep size distribution, and bluer KBOs have lower average albedos than red KBOs.
This could introduce an additional sampling bias, that would cause a bimodal population of small objects to appear primarily red.
The difficulty in understanding the inherent Cold object color distribution may be primarily a selection effect.

The application of combined datasets of KBO colors to understand inherent relationships between different KBO properties is a complicated process.
Understanding the selection effects and biases introduced by different survey procedures is difficult, if not impossible, and assumptions about these effects can influence the apparent acceptability of color distribution models.
The effects of KBO albedo may also result in an oversampling of red objects, and with the combination of these unknown effects we believe it is premature to claim conclusive trends in size-color relationships of KBOs.

\end{document}